\documentclass[12pt, a4paper]{article}
\usepackage{typearea}
\typearea{12}
\let\theta=\vartheta
\newcommand{\bi}[1]{\bibitem{#1}}
\def\be#1\ee{\begin{equation}#1\end{equation}}

 \newcommand\ph{\varphi} 
\renewcommand\o{\omega}

\begin{document}
\title{Gravity Probe C{\footnotesize lock} -- Probing the gravitomagnetic field
of the Earth by means of a clock experiment\footnote{To appear in the 
{\it Proceedings of the Alpbach Summer School 1997 ``Fundamental Physics in 
Space"}, organised by the Austrian and European Space Agency, A.~Wilson, ed.}}
\author{Frank Gronwald${}^\diamond$, Eleonora Gruber${}^*$, 
Herbert Lichtenegger${}^+$,\\ 
and Roland A. Puntigam${}^\diamond$ \\[1ex]
\small \it \\
\small\it ${}^\diamond$Institut  f\"ur Theoretische Physik\\ 
\small \it Universit\"at zu K\"oln\\
\small\it D-50923 K\"oln, Germany\\
\small \it \\
\small \it ${}^*$Institut f\"ur allgemeine Physik \\
\small \it Technische Universit\"at Wien \\
\small \it Wiedner Hauptstra{\ss}e 8--10 \\
\small \it A-1040 Wien, Austria\\
\small \it \\
\small \it${}^+$\"Osterreichische Akademie der Wissenschaften \\
\small \it Institut f\"ur Weltraumforschung \\
\small \it Inffeldgasse 12 \\
\small \it A-8010 Graz, Austria \\
\small \it \\
\small \it \\}
\date{}
\maketitle

{\footnotesize
\abstract{\noindent We outline a mission with the aim of directly detecting 
the gravitomagnetic field of the Earth. This mission is called Gravity Probe C.
Gravity Probe C(lock) is based on a recently discovered and surprisingly large 
gravitomagnetic clock effect. The main idea is to compare the proper time of 
two standard clocks in direct and retrograde orbits around the Earth. 
After one orbit the proper time difference of two such clocks is
predicted to be of the order of $2\times 10^{-7}$ s. The conceptual 
difficulty to perform Gravity Probe C is expected to be comparable to that of 
the Gravity Probe B--mission. }}
\bigskip

\section{Introduction}
Within our solar system relativistic gravity theories 
can be tested only in the 
weak--field limit. The simple reason for this is the absence of strongly 
gravitating objects in this part of our universe. It follows that relativistic
gravity experiments in the solar system produce only small corrections to the 
Newtonian theory which are, typically, difficult to measure.
  
In order to describe relativistic gravity on macroscopic scales we usually use
Einstein's (classical) general relativity. It is well--known that in 
a weak field approximation the structure of general relativity becomes 
formally analogous to that of Maxwell's special relativistic
electrodynamics (see Ref.\ \cite{step90}, for example). 
It is then possible to describe far fields, 
which are generated by isolated charge and current distributions, by means of 
multipole expansions. This circumstance is familiar 
from Maxwell's theory where the
corresponding multipole moments are divided into electric ones, due
to electric charge distributions, and magnetic ones, due to electric
current distributions. An analogous terminology is used in the linearized
Einstein theory where gravitoelectric and gravitomagnetic fields are 
introduced according to whether they are due to mass distributions or
mass current distributions, respectively. 
The standard tests of Einstein's general relativity, like perihelion 
precession of Mercury, light deflection of the Sun, gravitational
redshifts, and radar time delays, are explained by relativistic gravitoelectric
corrections. However, they yield no direct information on relativistic 
gravitomagnetic corrections.\footnote{This is true, at least, in the 
approximation used.}

In general relativity there are also ``classic'' gravitomagnetic effects.
They can be accounted for by what is commonly called the Lense--Thirring 
effect, i.e. the ``dragging of inertial frames'' by a spinning mass.
The Lense--Thirring effect was discovered in 1918~\cite{lens18} but turned out
to be hard to verify: It was calculated by Schiff in 1960~\cite{schi60} how 
to measure this effect by means of the precession of an orbiting gyroscope. 
To detect this precession is the main objective 
of the Gravity Probe B--mission which is awaited to
be launched by NASA at the end of 1999: Highly sensitive gyroscopes
will be carried in a drag free--satellite around the Earth and are
expected to measure directly the gravitomagnetic 
field of the Earth~\cite{keis97}. A recent study 
indicates a verification of the Lense--Thirring 
effect by means of the numerical evaluation of data which were obtained from 
laser--ranging observations of the satellites LAGEOS and LAGEOS II 
\cite{ciuf97}. Here the orbital planes of the satellites can be viewed as 
very large gyroscopes which are embedded in the gravitomagnetic field of the
earth. Then the predicted Lense--Thirring precession of the orbital planes 
is analogous to that of an orbiting gyroscope, and this is what was 
numerically evaluated.  

In this article we propose a mission, in the following called 
Gravity Probe C, which, similar to Gravity Probe B, 
is designed to directly measure the gravitomagnetic field of the Earth.
We intend to make use of a remarkable gravitomagnetic clock effect
that was recently pointed out by Cohen and Mashhoon \cite{mash93}: Consider 
the difference in proper time of two standard clocks, following identical 
orbits around a rotating body in direct and retrograde motion, respectively. 
After a proper azimuthal period this time 
difference turns out to be proportional 
to $J/M c^2$, with $J$ the angular momentum and $M$ the mass of the rotating 
body. In the case of two standard clocks orbiting the Earth this time 
difference amounts to the order of $2\times 10^{-7}$ s. It seems to 
be not well--known that a gravitomagnetic clock effect of this order exists. 
Therefore we will give a complete derivation of it in the next Section 2. To 
actually measure this effect is the objective of Gravity Probe C. 
An experimental realization of Gravity Probe C is expected to be 
technically demanding. This will be discussed in Section 3.

\section{A remarkable gravitomagnetic clock effect}
The exterior spacetime of a system with mass $M$ and specific angular 
momentum $a=J/M$ can be described by the Kerr geometry. 
The Kerr geometry is an {\it exact} solution of the vacuum field 
equation of general relativity. In Boyer--Lindquist coordinates 
$(t, r, \theta, \ph)$ the Kerr geometry takes the form
\be
ds^2\,=\, -dt^2 + \Sigma\Bigl({1\over \Delta}\, dr^2 + d\theta^2\Bigr)
+(r^2+a^2) \sin^2\theta\, d\varphi^2 + 2M{r\over \Sigma}(dt-a\sin^2\theta\,
d\varphi)^2\,. \label{Kerr}
\ee
Here we introduced the standard abbreviations $\Sigma:=r^2+a^2\cos^2\theta$ 
and $\Delta := r^2- 2Mr + a^2$. Except otherwise indicated we use units such 
that the gravitational constant $G$ and the velocity of light $c$ are set to 
unity, $G=c=1$. The following derivation of the gravitomagnetic clock
effect under consideration is based on Ref.\ \cite{mash93}.

What we first want to calculate is the proper time as shown by a standard 
clock which follows a geodesic in the Kerr geometry. To keep things
simple we specialize on a circular, geodesic orbit, i.e., we put 
$r={\rm const}$ and $\theta = {\pi\over 2}$. The 
geodesic equation which interrelates the remaining coordinates $t$ and $\ph$
is given by 
\be
{{d^2 r}\over{ d\tau^2}} + \Gamma^r_{ij} {{dx^i}\over{d\tau}}
{{dx^j}\over{d\tau}}\,=\, \Gamma^r_{ij} {{dx^i}\over{d\tau}}
{{dx^j}\over{d\tau}}\,=\,0\,. \label{geodesic}
\ee
(We note the use of the 
Einstein summation convention which involves the indices 
$i$ and $j$.) The Christoffel symbols $\Gamma^r_{ij}$ are determined 
from the metric (\ref{Kerr}). 
The geodesic equation (\ref{geodesic}) turns out to be
\be
dt^2-2a\, d\varphi dt + \Bigl(a^2 - {{r^3}\over M}\Bigl) d\varphi^2 \,=\, 0\,.
\ee
This quadratic equation can be solved immediately. The result is
\be
{{dt}\over{d\varphi}}\, = \,a\pm\Bigl({{r^3}\over{M}}\Bigl)^{1\over 2}
 \,=:\,a\pm {1\over{\omega_0}}\,. \label{result}
\ee
Here we introduced the symbol $\omega_0$ to denote the Kepler 
(and Schwarzschild!) result 
\be
{{d\ph}\over{dt}} = \pm \Bigl({M\over{r^3}}\Bigr)^{1\over 2} =:\omega_0\,.
\ee

The square of a proper time interval $d\tau^2$ is given, up to a factor 
$c^2$ and a sign, by the line element (\ref{Kerr}). For a circular,
equatorial orbit we find from (\ref{Kerr}) the relation
\be
\Bigl({{d\tau}\over{d\ph}}\Bigr)^2 \,=\, 
\Bigl(1-{{2M}\over{r}}\Bigl)\, \Bigl({{d t}\over{d\ph}}\Bigr)^2
+ {{4Ma}\over{r}}\, {{dt}\over{d\ph}}
-r^2 -a^2\Bigl(1+{{2M}\over{r}}\Bigr)\,.
\ee
We substitute (\ref{result}) and obtain
\be
\Bigl({{d\tau}\over{d\ph}}\Bigr)^2 \,=\, {1\over{{\omega_0}^2}}
\Bigl(1-{{3M}\over r} \pm 2a\o_0\Bigr)\,,
\ee
or
\be
{{d\tau}\over{d\ph}}\,=\, 
\pm{{1}\over{\o_0}}\sqrt{1-{{3M}\over r}\pm 2a\o_0}\,.
\ee
Plus and minus sign apply to direct and retrograde orbits, respectively.
For one closed orbit $(\ph\longrightarrow \ph + 2\pi)$ it is now trivial to
obtain the integrated proper time as
\be
\tau_\pm \,=\, {{2\pi}\over{\o_0}} \sqrt{1-{{3M}\over r}\pm 2a\o_0}\,.
\label{exact}
\ee
This result is exact. 

We are interested in the proper time difference
$\tau_+ - \tau_-$. We Taylor-expand the square root of (\ref{exact}) 
in powers of $a$ and obtain to first order
\be
\tau_+ - \tau_- \, \approx \, 4\pi a \,=\, 4\pi {{J}\over{M}}\,,
\ee
or, in natural units,
\be
\tau_+ - \tau_- \,= \, 4\pi{{J}\over{M c^2}} +  {\cal O}(c^{-4}) + ...\,.
\ee

This formula manifests the gravitomagnetic clock effect we are interested
in. It is interesting to note that, in this approximation,
the difference $ \tau_+ - \tau_-$ is independent of the radius $r$
of the orbit and
the gravitational constant $G$.\footnote{This observation 
is discussed in more detail
in Ref.\ \cite{mash97}.} In case of orbits around the Earth we put
$M\approx 6\times 10^{24}\, {\rm kg}$, $J\approx 10^{34}\,{\rm kg}\,
{\rm m}^2 \, {\rm s}^{-1}$, and find
\be
\tau_+ - \tau_- \, \approx \, 2\times 10^{-7} {\rm s}\,.
\label{oureffect}
\ee
This time difference is surprisingly large. Time shifts due to the 
gravitomagnetic field of the Earth are widely believed to be of much lower 
order. Indeed, suppose the time difference of a direct and retrograde moving
clock is taken at a {\it fixed time}, say, after one Kepler period
$T_0=2\pi /\omega_0$. Then it is straightforward to show that for
a circular equatorial orbit,
\be
\tau_+ - \tau_- \, = \, 12\pi\frac{GJ}{c^4r} + {\cal O}(c^{-6}) + ... \;.
\ee
As an example, we set $r\approx 7000\, {\rm km}$ and obtain 
\be
\tau_+ - \tau_- \, \approx \, 3\times 10^{-16} {\rm s}\,.
\ee
This is nine orders of magnitude smaller than the time difference 
(\ref{oureffect}). We recall that the result (\ref{oureffect})
presupposes that the time difference of the two clocks is taken with respect
to a {\it fixed angle $\varphi$} (i.e., after each clock has covered an 
azimuthal interval of $2\pi$) and not with respect to a fixed 
time. It is very important to note this conceptual difference in order to
avoid confusion. 

The gravitomagnetic clock effect of 
this section can be generalized to circular orbits with 
a nonvanishing inclination to the equatorial plane,
i.e., to circular orbits which are not restricted to $\theta = {\pi \over 2}$
\cite{mash97, thei85}. It turns out that with increasing inclination
the time difference $\tau_+ - \tau_-$ becomes smaller and finally vanishes
in case of a polar orbit. Analytic calculations of $\tau_+ - \tau_-$ 
for an arbitrary (eccentric) orbit have not been conducted, yet. It is
expected that the corresponding geodesic equations can only be solved 
by means of numerical integration. 

\section{Realization of Gravity Probe C}

In principle, it is a trivial task to measure a time 
difference of $2\times 10^{-7}{\rm s}$ with today's technology. 
However, an experimental 
verification of the gravitomagnetic clock effect does not only require 
the measurement of the time difference between 
two well--defined events up to an
accuracy of $2\times 10^{-7}$ s. It is, in fact, the proper time along the 
direct and retrograde orbits that is used as a clock. 
Therefore, it is not sufficient to send two highly accurate and stable clocks 
into space and let them orbit in opposite directions. The orbits 
themselves have to be highly accurate and stable as well. We recall that 
in order to obtain the time difference (\ref{oureffect}) we have to 
subtract two periods, each of which represents 
the sum of a Kepler period and a much smaller relativistic correction. 
Under the assumption of {\it identical} orbits, the Kepler periods and their
gravitoelectric corrections cancel upon subtraction while the gravitomagnetic
contributions add up, yielding the actual clock effect under consideration
here. Disturbances 
of the orbits will in general change the Kepler periods  
of the orbiting clocks. It follows that in this case the Kepler periods will 
not exactly cancel but may exhibit a significant time difference.

For an (equatorial and circular) orbit of Keplerian period $T_0$,
the relative gravitomagnetic variation of the orbital period is given by
\be
       \frac{\tau_+ - \tau_-}{T_0}\,=\, \frac{2\,J}{M\,c^2}
       \left(\frac{G\,M}{r^3}\right)^{1/2}   \, .
\label{relative}
\ee
In case of a near-Earth orbit
this expression amounts to about $4\times 10^{-11}$.
It was noted in \cite{mash93} that this expression is exactly the same as the 
the relative gra\-vito\-magnetic precession angle of a Gravity 
Probe B--gyroscope:
The precession angle $\Omega_P$ (per $2\pi$ radians) of a 
gyroscope after one orbit is predicted to be
\be
\Omega_P \,=\, \frac{2\,G\,J}{c^2\,r^3}\,,
\ee
such that
\be
   \frac{\Omega_P}{\omega_0} \,=\, \frac{2\,G\,J}{c^2\,r^3}
   \left(\frac{G\,M}{r^3}\right)^{-1/2}  \, .
\ee
This is identical to equation (\ref{relative}).
Therefore it is expected that the difficulty of performing the Gravity Probe C 
mission is essentially equivalent to that of Gravity Probe~B.

\subsection{Mission design}
Gravity Probe C is simple from a conceptual point of view. To perform Gravity 
Probe~C requires to collect precise data from direct and retrograde orbits 
around the Earth. It follows from the previous discussion of the 
gravitomagnetic clock effect that these orbits should be as circular and as
equatorial as possible. In order to collect the data required one could 
put atomic clocks on board of satellites and send them on specific 
orbits.

One should keep in mind that future space missions are expected to carry 
highly stable and accurate clocks \cite{solo97}. 
Also there already exist satellite--based
measuring--devices, like the Global Positioning System (GPS), which make
it possible to determine precisely the positions of orbiting clocks in space.
Therefore, Gravity Probe C could rely, in principle, on data which are
collected in the context of other space missions. 

In order to become more 
specific and to obtain some error budget we now assume a realization
of Gravity Probe C in the form of atomic clocks on board of satellites and
in direct and retrograde motion around the Earth.
We divide the error sources of such an experiment in two groups, namely
\begin{itemize}
\item[(i)] errors due to the tracking of the actual orbits, and
\item[(ii)] deviations from idealized orbits due to
\begin{itemize}
\item{} mass multipole moments of the Earth
\item{} radiation pressure
\item{} gravitational influence of the Moon, the Sun, and other planets
\item{} other systematic errors (e.g., atmospheric disturbances).
\end{itemize}
\end{itemize}

\subsection{Errors due to the tracking of orbits}
The tracking of the actual orbits requires the measurement of distances 
and angles. (We note in passing that the angle $\varphi$ 
has to be measured with respect to
a fixed star.) The position of a satellite along an orbit can be determined to
a few centimeters using a Global Positioning System; therefore, the
temporal uncertainty that a near-Earth satellite has actually returned to the
same azimuthal position in space can be roughly estimated to be 
$\delta\,\tau \sim \delta\,r/v \sim 10^{-6}\,{\rm s}$. Here
$\delta\,r \sim 1\,{\rm cm}$ is the position uncertainty along track and $v$ is
the orbital speed of the satellite. The gravitomagnetic clock effect, however,
involves a definite temporal deviation of $10^{-7}\,{\rm s}$. 
A simple but more detailed error estimate, based on the formulas provided
in Section 2, shows that, to first order in uncertainties $\Delta \varphi$ and
$\Delta r$,
\be
\tau_+ - \tau_- \,=\, 4\pi{{J}\over{Mc^2}} + \sqrt{{r^3}\over{GM}}\Delta\varphi
+ 3\pi\sqrt{{r}\over{GM}}\Delta r \,.
\ee
From this it is easy to see that one should be able to measure the 
orbital radius up to an accuracy of the order of $10^{-4}\,$m and to 
determine angles up to an accuracy of $10^{-10}\,$rad in order to 
keep the errors due to the measurement smaller
than the clock effect after one orbit. These requirements are about one order
of magnitude higher than what can be achieved with today's technology. 
However, one should keep in mind that the clock effect 
is cumulative -- just like the precession angle of a GP-B gyroscope -- and 
hence many orbital periods can be used for a measurement of the 
gravitomagnetic effect; that is, the statistical tracking errors could be
overcome if one were able to perform many single measurements.

\subsection{Systematic errors}
The systematic errors of the second group of error sources 
have a more serious influence 
on the gravitomagnetic clock effect. In order to calculate the influence 
of such perturbative accelerations on the Kepler period one has to 
focus on sections of orbits rather than on complete closed 
orbits. This is because under favorable conditions (e.g., an almost 
constant perturbative acceleration) different temporal deviations can cancel 
each other when summed over a closed orbit. 
It is possible to show from Newtonian mechanics that perturbative 
accelerations should be kept below $10^{-11}\,$g in order for the clock 
effect to become measurable.

\subsubsection{Influence of multipole moments of the Earth}
Perturbative accelerations due to multipole moments of the Earth are of the 
order of $10^{-3}\,$g. That is, the orbit of a satellite under the influence 
of the nonspherical and inhomogeneous form of the Earth resembles something
like a bumpy road. However, we do know the gravitational field of the Earth 
up to an accuracy of $10^{-9}\,{\rm g} - 10^{-10}\,$g. 
NASA's already approved gravity mapping mission GRACE  
is expected to push this accuracy higher by about two orders of magnitude.  
This would then make it possible, in principle, to correct 
for the influence of the multipole moments of the Earth 
on a gravitomagnetic clock experiment. 

\subsubsection{Influence of radiation pressure} 
The radiation pressure of the Sun causes perturbative accelerations of
the order of $10^{-8}\,$g. Using drag--free satellite techniques, this
disturbance can be downsized by two orders of magnitude to $10^{-10}\,$g.
To keep this error source under control one thus has to be able
to determine the solar pressure accurately enough such that corrective
calculations can be performed. Alternatively, one must wait 
for drag--free satellites that
can perform at least one order of magnitude better than current technology.

\subsubsection{Influence of the gravitational fields of the Sun and other 
planets}
The gravitational fields of the Moon 
and the Sun cause relative accelerations  
between the Earth and the orbiting clocks. The amplitudes
of these accelerations are of the order of
$10^{-7}\,$g (Moon) and $10^{-8}\,$g (Sun). The influence of the other
planets of the solar system plays only a minor role. For Jupiter, e.g.,
we obtain an influence of the order of $10^{-12}\,$g. The positions of the
Moon and the Sun are known with much higher accuracy than is needed to 
determine their gravitational field at the level of $10^{-11}\,$g;
therefore, in principle, the influence of the Moon and the Sun can be 
properly taken into account.

\subsection{Concluding remarks}
It is clear from the considerations of this section that Gravity Probe C 
requires high precision measurements. Overall, the accuracy required 
cannot be reached with today's technology; however, what is missing 
is a factor of the order of 10. As technology advances in time, a 
gravitomagnetic clock experiment might become 
realizable soon.\footnote{One should
remember that it took more than 35 years to realize the Gravity
Probe B mission.} With higher accuracy available the data required
for Gravity Probe C might be obtained as a by--product of other 
space missions. 

Gravity Probe C will be kept in mind. 
This is also true for the gravitomagnetic clock 
effect of this paper which might also become
important in the context of other tests of general relativity. 

\section*{Acknowledgments}
We would like to thank the organizers of the Alpbach Summer School 1997 
``Fundamental Physics in Space" for the highly interesting and pleasant
meeting. Financial support from ASA and DARA was greatly appreciated.
Thanks are also due to all participants which showed us their interest
in Gravity Probe C. In particular we would like to acknowledge 
enlightening and helpful discussions with P.L.\ Bender, D.B.\ DeBra, 
R.\ Laurance, G.M.\ Keiser, M.D.\ Moura and, last not least, G.\ Sch\"afer. 
The physics behind Gravity Probe C was kindly shared with us by
B.\ Mashhoon.

\end{document}